\documentclass[preprint,showpacs,preprintnumbers,amsmath,amssymb]{revtex4}

\usepackage{graphicx}% Include figure files
\usepackage{dcolumn}% Align table columns on decimal point
\usepackage{bm}% bold math
\usepackage{doublespace}
\begin{document}
\noindent \textbf{Reply to Comment on "Microscopic theory of
network
glasses"}\\

In the preceding Comment\cite{Micoulaut2003}, Micoulaut and
Boolchand (MB) compare our predictions\cite{HallWolynes2003} on
glass transitions in network systems for the ratios of the
transition temperatures T$_A $/T$_G $ and T$_K $/T$_G $ with
experimental data for Ge$_x $Se$_{1-x}$. T$_A $ is the temperature
at which "landscape dominated" activated dynamics begins and T$_K
$ is the ideal glass temperature where an entropy crisis occurs.
T$_G $ is the temperature where the kinetic glass transition
occurs in the laboratory. MB claim significant differences between
experiment and theory are seen and claim the lack of agreement is
due to the failure of our theory to include "local structures,
structural self-organization, and non-central forces".

The T$_K $/T$_G $ is ratio shown in Fig. 1 of \cite{Micoulaut2003}
and obtained from Angell's group\cite{Tatsumisago1990}.
Examination of Fig. 1 in fact shows remarkable agreement between
experiment and theory. Not only is the decrease in T$_K $/T$_G $
with n$_b $ (the average number of bonds/atom) reproduced by the
theory but the difference between experiment and theory is at most
10\%. In light of the admitted simplifications in the theory (as
discussed below), the comparison lends strong support to the
theory. Our theory did not predict the detailed dynamics beyond
the degree of bonding where rubber behavior ensues but this
"overconstrained" behavior is precisely what Angell claims to
begin for n$_b
> 2.4$ in the Ge$_x $Se$_{1-x}$ system and is claimed to be the origin of the
minimum discussed by MB. Quantitatively, our n$_b $ rigidity limit
is n$_b $ = 2.9.

On the other hand, the comparison of T$_A $/T$_G $ ratios made by
MB is misleading because the T$_A $ defined by them is not the
T$_A $ defined in the random first order transition theory of
glasses. MB define T$_A $ as the temperature at which non-ergodic
behavior in $\stackrel{.}{H}$ is not apparent. Their lack of
non-ergodicity is equated to a lack of activated behavior. Yet
whether non-ergodicity is seen depends on the time scale of
scanning in the experiment and is thus not a landscape parameter
like T$_A $ which does not depend on the measurement time scale
but rather signals the origin of activated behavior with rate
slower than local vibrations. The temperature at which the
viscosity of Ge$_x $Se$_{1-x}$ begins to show activated behavior
must be considerably greater than 10\% higher than T$_G $, as
would be suggested by the plots in the preceding Comment.
Lubchenko and Wolynes\cite{Wolynes2003} show that T$_A $ occurs
for most substances at a temperature higher than that at which the
viscosity is 100 poise. Extrapolating the Angell viscosity
data\cite{Tatsumisago1990} of either the ternary mixture GeAsSe or
GeSe (inset b of Fig 2 in Ref.\cite{Tatsumisago1990}) to 100
poise, gives a T$_A $ at least 1.7 times T$_G $. Nevertheless we
agree and pointed out in out Letter that the predicted T$_A$/T$_G$
for soft repulsive spheres with fixed harmonic connections is too
large for real molecular systems. Density fluctuations, the
breakdown of fixed bonding relations as the boiling point is
approached, and attractive forces will clearly affect the
predictions of our theory (which included only a soft repulsion),
as discussed in our Letter. Our prediction is thus merely saying
the entire liquid regime encompasses activated behavior for
substances with sufficient bonding. We point out that Angell
suggests that liquids become "stronger" as n$_b $ is increased,
again just as our theory predicts.

MB suggest several features that are lacking from our theory. Two
of these, local structure (by which we assume MB mean lack of
complete chemical mixing) and non-central forces (bond-bending),
were explicitly mentioned by us, but not included in the
calculations so as to highlight the basic physics. These can be
easily included. "Structural self-organization" by which MB
apparently mean large scale structure formation in the network was
not discussed in our Letter. Phenomena of this type
("polyamorphism") seem to occur in several glassy fluids. They may
be essential to the glass transition. On the other hand, in our
view, it is likely that these are  consequences of the low
configurational entropy density near T$_G $ which allows weak
forces to cause large scale ordering. There is no reason to
believe the glass transition is the only transition possible in a
fluid phase and we do not mean to imply that it is.

In summary, the T$_K $/T$_G $ comparison made by MB  provides
strong evidence for the validity of the random first order
transition theory of network liquids, while the T$_A $/T$_G $
comparison is based on the misidentification of T$_A$ with a
nonequilibrium scan time scale dependent quality and is not valid.
Estimating T$_A $/T$_G $ from viscosity data provides evidence
that the trends of how T$_A $/T$_G $ varies with bonding are
indeed reproduced by our theory but quantitative improvements are
needed. Extensions of the theory to include the effects of
attractive forces, bond-bending and complex local structure
formation are not hard to develop.

R.W. Hall$^1 $ and P.G. Wolynes$^2 $

\noindent $^1 $ Department of Chemistry, Louisiana State
University, Baton Rouge, Louisiana 70803-1804\\
\noindent $^2 $ Department of Chemistry and Biochemistry,
University of California, San Diego, La Jolla, California
92093-0332\\

\bibliography{networkglass}
\end{document}